# Methodological thoughts on expected loss estimation for IFRS 9 impairment: hidden reserves, cyclical loss predictions and LGD backtesting

Wolfgang Reitgruber[1][2]

**Abstract**

After the release of the final accounting standards for impairment in July 2014 by the IASB, banks will face the next significant methodological challenge after Basel 2. In this paper, first methodological thoughts are presented, and ways how to approach underlying questions are proposed.

It starts with a detailed discussion of the structural conservatism in the final standard. The exposure value iACV$^{©}$ (idealized Amortized Cost Value), as originally introduced in the Exposure Draft 2009 (ED 2009), will be interpreted as economic value under amortized cost accounting and provides the valuation benchmark under IFRS 9. Consequently, iACV$^{©}$ can be used to quantify conservatism (ie potential hidden reserves) in the actual implementation of the final standard and to separate operational side-effects caused by the local implementation from actual credit risk impacts.

The second part continues with a quantification of expected credit losses based on "Impact of Risk$^{©}$" instead of traditional cost of risk measures. An objective framework is suggested which allows for improved testing of forward looking credit risk estimates during credit cycles. This framework will prove useful to mitigate overly pro-cyclical provisioning and to reduce earnings volatility.

Finally, an LGD monitoring and backtesting approach, applicable under regulatory requirements and accounting standards as well, is proposed. On basis of the "NPL Dashboard", part of the "Impact of Risk$^{©}$" framework, specific key risk indicators are introduced that allow for a detailed assessment of collections performance versus LGD in in NPL portfolio (bucket 3).

---

[1] Department for Strategic Risk Management & Control, UniCredit Bank Austria AG,
Julius Tandlerplatz 3, A-1090 Vienna, Austria; email: wolfgang.reitgruber@unicreditgroup.at

[2] The presented opinions and methods in this paper are solely the responsibility of the author and should not be interpreted as reflecting those of UniCredit Bank Austria AG or of the author's former employers. The concepts and terms "Impact of Risk", "IoR", "iACV", "PL Dashboard" and "NPL Dashboard" are copyright by the author, all rights reserved. For commercial, profit-oriented or capital-optimizing applications of these concepts (either within a corporation or for consulting/training purpose), written permission by the author is required.

**1. Introduction**

After many years of preparation, development of drafts and feedback sessions, the final standard of IFRS 9 impairment was released in July 2014. The change from an incurred loss to an expected loss model for loan loss provisioning went into its final round. Methodological and stochastic concepts of credit risk, so far mostly used in internal steering, in regulatory context for IRBA segments and in modeling economic capital, will enter the area of accounting.

The timing seems just right: during the past couple of years since introduction of IRBA concepts with Basel 2, the banking industry developed high quality credit risk models for regulatory capital under pillar 1 as well as for economic capital under pillar 2. Knowhow in development and validation units was created and the use and acceptance of credit risk methods in actual risk management processes significantly increased.

The expectations on the quality of credit risk parameters under the final standard of IFRS 9 impairment (short: the "final standard") are at least as serious as for IRBA: Loan loss provisions need to be based on unbiased and objectively justified methods, considering all reasonable information concerning past and present information, and future expectations (forward looking).

Only the actual implementation and the interpretation of the final standard by financial institutions and auditors will prove how far these expectations can be met in real life. Even the standard setters themselves are aware of this uncertainty: "... *cannot quantify the magnitude of the impact of moving to the new impairment model on an entity's financial reporting ... While all entities will be required to meet the objective of the impairment requirements ..., in practice, the loss allowance will depend in part on how entities operationalize IFRS 9*".

This paper primarily refers to the final standard on impairment, as released in July 2014 by the IASB. The proposed methods or interpretations are partly hypothetical and may have to be adjusted depending on how certain rules are finally implemented or interpreted. Although all suggestions were developed in good faith, the author does not assume any further responsibility. Before implementing the proposed methods, any relevant updates or current interpretations of the standard have to be considered adequately.

After a short introduction, the analytical part of this paper will focus on the following topics:

- The final standard as approximation of a methodologically clean approach. It will be shown, how the valuation method introduced with the ED 2009 provides an unbiased, fair and generic benchmark iACV$^{©}$. Unbiased and fair because it can be interpreted as fair value under idealized market conditions. Generic because it neither requires specific risk processes such as identification of loss events for bucket 3 or risk increase events for bucket 2, nor does it require ad-hoc measures such as the 12-month expected loss provision in bucket 1. iACV$^©$ provides a continuous quantification of credit risk without cliff effects.

- By developing, proposing and discussing intermediate drafts before finalizing the framework, the IASB tried to propose operationally simpler approaches without getting overly conservative. The quantification of potential hidden reserves, introduced in this process, will be discussed in the following section. The ongoing monitoring of this amount should be part of an efficient IFRS 9 reporting. Additionally it provides a basis to differentiate between process impacts caused by local choices in implementing the final standard eg how the identification of risk increase is actually implemented, and real portfolio developments in terms of credit risk.

   Remark 1.1: The implementation of ED 2009 in accounting systems was rejected because of operational effort and complexity: it would have required full alignment of risk and finance databases and valuation methods. In the suggested analytical framework however, with reduced data requirements regarding quality and timeliness, an analytical (offline) implementation seems reasonable and justified by the gain in transparency it provides.

- The need to consider all reasonably available information – especially cyclical, forward looking predictions – in producing loss estimates will become another analytical challenge. Credit risk does exhibit high volatility, which makes the objective quantification and verification of cyclical effects difficult. Traditional key risk indicators such as loan loss provisions or cost of risk cannot easily be used for analytical purpose: they are observed mostly annually (ie resulting time series are too short for meaningful statistical analysis), and may be distorted because of strong management attention on profit and loss (P&L). On the other hand, economic predictions themselves are not very reliable, especially concerning the proper timing and depth of an anticipated crisis. Additionally the impact of any specific crisis may significantly differ by industry and business segment.

   The framework "Impact of Risk$^©$", as introduced in Reitgruber (2013), provides a high-frequent (monthly) and objective measurement of credit risk. It allows for statistically viable testing and quantification of historical crisis situations with respect to their impact on credit risk for specific portfolio segments. The implementation of these tests to verify the quality of predictions is necessary to avoid overly cyclical loss estimates and to mitigate the potential risk increase in the sector because of increasing earnings volatility.

- Finally, the question of provision amounts for the defaulted portfolio (bucket 3) is addressed. The final standard states that already existing (ie regulatory) parameters should be re-used as far as possible. For the defaulted portfolio this statements primarily relates to LGD estimates for the NPL portfolio and material values of collateral modeled through haircut estimates. On the other hand, specific and explicit requirements are imposed for impairment accounting: Estimates need to be unbiased and derived on basis of a probability-weighted expectation of discounted cash flows. This is mostly comparable to regulatory requirements. Additionally, the effective interest rate has to be used as discount rate and forward looking expectations regarding the future development of the portfolio have to be considered. These requirements seem to have little in common with regulatory standards. A monitoring and validation technique based on "Impact of Risk$^©$" is presented, which allows testing a given LGD method simultaneously against regulatory and accounting standards. The underlying quantitative indicators will prove useful in monitoring collections efficiency.

The presented tools and methods will provide valuable insight during the implementation phase of the final standard. After rollout of the new accounting standards, these key risk indicators will continue to improve the understanding of portfolio dynamics on an ongoing basis when applied in active portfolio monitoring. Overall, the use of these measures for credit risk portfolios before and after default will have a positive contribution to the general understanding of credit risk related processes and their efficiency.

This paper was inspired by the risk philosophy outlined in Taleb (2012) and the terminology of the known, the unknown and the unknowable as summarized in Diebold et al (2010). Financial institutions will have to deepen their understanding of the stochastic properties of day-to-day credit risk and related business processes. The supporting role of capital as cushion for small and large credit risk shocks is crucial to develop proper systematic responses for these events, to improve internal processes on basis of objective quality standards. The framework "Impact of Risk$^©$" and its modules "PL Dashboard" and "NPL Dashboard" will move credit risk from the "unknown" or even "unknowable" at least one step closer to the "known".

**2. The methodological framework**

The entire paper covers accounting of financial instruments under amortized cost, based on the effective interest method. Financial instruments are to be reported this way under IFRS 9, if contractual cash flows consist primarily of interest and principal payments and the business model aims at collecting these cash flows (relevant for most relevant vanilla type loan portfolios).

Although the final standard and IRBA regulations share similar concepts to measure credit risk (Expected loss modelled through PD, EAD and LGD), they pursue different objectives: Main focus of impairment accounting is the expected loss, whereas regulatory standards in pillar 1 or 2 focus on capital requirements to cover unexpected loss. The uncertainty (volatility) of future cash flows, main driver for fair value considerations, is of little relevance for amortized cost accounting.

Frequency and volatility of measurements: credit risk is typically reported once a year in high quality – the operational effort to ensure daily updates with sufficient quality is significant. Reasonably long time series without structural breaks to backtest portfolio-loss models are simply not available. High volatility due to its stochastic properties (rare, significant events with complex dependence structures) makes an empirical analysis even more challenging.

Strategic objective: Management tends to use cost of risk in order to achieve different strategic goals – smoothing the P&L, displaying financial health ("Signaling") or performing portfolio cleanup, if budgets or market expectations cannot be achieved ("Big Bath Accounting"). An overview of the different managerial strategies can be found in Gaber (2013). In Domikowsky et al (2014) the situation in Germany is empirically analyzed, where certain types of impairment are used for tax optimization. Because of these distortions, the empirical analysis based on cost of risk is challenging.

The methodological link from provisioning to capital requirements under the IRBA is represented by the shortfall, defined by the difference between the expected loss and the provision amount. For capital requirements, this leads to an effective replacement of the provision amount by a parameter-based assessment of the expected credit risk (the expected loss). However, changes in capital requirements caused by the shortfall receive much lower management attention than changes to provisions. For P&L, relatively small changes already receive high attention. For capital however, the absolute magnitude is more relevant. Significant large changes are mostly driven by external or regulatory factors. Comparably small monthly changes, triggered by portfolio developments and shortfall adjustments, tend to receive less attention. This situation will change with IFRS 9, when these changes are going to hit the P&L immediately. The underlying dynamic of monthly capital requirements and P&L impacts under IFRS 9 caused by expected loss changes is quantified in the "Impact of Risk" methodology proposed by Reitgruber (2013). A short introduction to this framework is provided in Appendix B.

The transition from IAS 39 to IFRS 9 in provisioning will result in an unprecedented direct influence of statistical methods (the parameters PD, LGD and EAD) to the P&L. In this sense, the methodological concepts supporting the final standard have to be seen as an extension of the methodological framework already introduced for IRBA portfolios under Basel 2.

The increase in provisioning levels, to be realized with the introduction of the final standard caused primarily by the extension of the forecast period to one year or lifetime, will be significant. These amounts will most likely affect certain methodological decisions in the preparation phase. In the context of the ongoing application however, other effects will prevail: Reduced flexibility in the provisioning process may cause an increase in the volatility of monthly, quarterly or annual reported risk costs. This increase in volatility is already addressed in empirical studies, eg in Grünberger (2012). Although this effect should not surprise analysts, it may have unexpected consequences in the context of disclosure and market perception. Deeper methodological understanding of the volatility of credit risk has to be developed to provide adequate mitigation.

We can conclude the following: Currently, P&L impacts and capital requirements are seen as largely separate disciplines. This situation will definitely change with the implementation of the final standard. The increase in provisions volatility will lead to a reduction of predictability of reported credit risk. On the other hand, fluctuations will have to be compensated based on existing regulatory capital, fulfilling its main purpose. The increased volatility of results in a practically relevant 5 to 20-year observation period represents a major challenge in the management of financial institutions. Focus will shift from the 99.9x% events under "gone concern" scenario used for economic and regulatory capital to the lower 80% to 95% quantiles of the loss distribution for day-to-day "going concern" analysis.

**3. Introducing the benchmark iACV©: economic value under amortized cost accouting**

For a discussion of lifetime risk concepts the following technical terms need to be introduced:

Let $GCA_t$ be the gross carrying amount according to the final standard at time t. It is calculated on basis of <u>contractually agreed</u> cash flows $CF_t$, discounted by the applicable effective interest rate i. The net carrying amount ($NCA_t$) is derived from the $GCA_t$ by deduction of loan loss provisions. Provisions are calculated as the sum of expected losses $R_t$, discounted by the effective interest rate i.

Let $iACV_t$ be the exposure value according to ED 2009 at time t. In contrast to $GCA_t$ it is based on <u>expected</u> cash flows $CF_t - R_t$ and $i_{ED}$ as applicable discount rate. The discount rates in both cases are implicitly defined such that both $GCA_0$ and $iACV_0$ are identical to the initial loan amount net of fees according to B5.4.1ff from the application guidance in the final standard.

The risk level $r \coloneqq i - i_{ED}$ is derived as difference between these discount rates. This definition of risk level corresponds to the common understanding of a risk margin. It is a measure of the average (lifetime) expected loss, given as a percentage of $GCA_t$. The uncertainty of future cash flows, as measured in the unexpected loss, has no impact on this risk level.

The risk profile $p_t := \frac{R_t}{r\, GCA_t}$ describes the temporal distribution of the expected credit losses over the life of the loan, putting loss amounts in relation to risk level r and $GCA_t$. The neutral risk profile is characterized by flat 100%, which means that losses of size $r\, GCA_t$ are expected at any time. This neutral risk profile is conceptually similar to a situation with constant conditional forward PDs (ie constant hazard rates).

The economic value of a financial asset is generally defined as the sum of (expected) future cash flows, discounted at an interest rate to properly asses the time value of money. iACV$^©$ fulfils this condition: Expected cash flows correspond to contractual cash flows net of expected cash shortfalls = losses are the basis. The discount rate is implicitly defined to align the initial loan amount (net of applicable fees) with discounted expected future cash flows. The resulting discount rate consequently reflects the time value of money, net of risk costs as mentioned before.

Changes in the market situation – besides changes to expected losses – shall have no effect on the valuation of a financial asset at amortized cost. The underlying methodological assumption is the stability of market conditions. This assumption has to cover the cost and funding structure (ie operational cost, expectations of risk level and volatility, profit expectations, etc), and a flat yield curve to ensure stability of the applicable discount rate even under decreasing residual maturity.

To estimate iACV$^©$ the discount rate at loan origination remains fixed over the entire life of the financial instrument – in line with the assumptions for a stable market environment as defined above. In the ongoing calculation of $iACV_t$, expected future cash flows are updated for new loss expectations. Consequently, the $iACV_t$ describes the economic value for all t for stable market conditions, only adjusted for changed (lifetime) loss expectations.

Ultimately, the final standard also emphasizes this fact: "*In the IASB's view, expected credit losses are most faithfully represented by the proposals in the 2009 Impairment Exposure Draft…*".

Consequently, differences between $iACV_t$ and $NCA_t$ are not representing portfolio risk characteristics but quantify the effect of specific operational details in the implementation of the final standard. Main drivers for these differences are the processes and thresholds defining the transition between buckets 1, 2 and 3 and certain conservative choices in the final standard such as the day-one provision of 12-month expected loss.

The key argument against implementation of the ED 2009 was finally its operative complexity. Finance and risk systems must be fully aligned and mutual dependency is created. Cost-benefit considerations led to alternative models, which seemed easier to implement, but with the tendency towards more conservative exposure valuations.

Summarizing, $iACV_t$ represents the appropriate methodological benchmark against which the calibration (or conservatism) of the local implementation of the final standards can be compared to. As a result, the discussion of portfolio movements can be split into two components: the quantification of the difference between $NCA_t$ based on the final standard and $iACV_t$, and actual credit risk portfolio dynamics, measured by $iACV_t$.

### 4. $NCA_t$ compared to $iACV_t$: the case of unchanged risk level

In contrast to $iACV_t$, the gross carrying amount $GCA_t$ in the final standard is not sensitive with respect to changes in risk level or risk profiles with non-constant hazard rates. $GCA_t$ is based purely on contractual details and considers risk expectations only through the initially applied effective interest rate. At loan origination, both standards lead to identical exposure values $iACV_0 = GCA_0$ by definition, although different cash flows and different discount rates are applied. In case of neutral non-varying risk profile (constant hazard rate) and unchanged risk level this relationship holds true until end of maturity. The proof of this statement can be found in Lemma A 1 in Appendix A: $iACV_t = GCA_t$ under the condition of a constant hazard rate. In this situation $iACV_t - NCA_t = iACV_t - GCA_t + EL = EL > 0$ proves that the final standard is strictly conservative (by EL = the 12-months expected loss) compared to the economic value under stable market conditions and constant hazard rates (in bucket 1).

Unfortunately neutral risk profiles are not the typical case. The empirical analysis of credit risk frequently shows that default events tend to occur at later stages, with a delay. Bullet loans are explicitly mentioned in the final standard, where low ongoing installments may cause reduced default rates in the beginning. Defaults are finally triggered by the bullet payment, leading to higher default rates towards end of maturity. A similar effect is plausible for other portfolios as well, where artificially low default rates prevail in the first year (eg revolving loans, long-term contracts).

Risks, which are more relevant right after loan origination (eg fraud risk) are generally controlled more efficiently in developed loan portfolios. They could provide some compensation in terms of loss distribution in the other direction. For simplicity however, we will focus in the following discussion on the case of delayed risk occurrence.

An immediate consequence of this delay is that $GCA_t$ in the final standard tends to become more aggressive than $iACV_t$. A delay of about one year leads to a difference corresponding to the 12-month expected loss. Longer delays may lead to even higher deviations. Please find a detailed quantitative proof in Corollary A 3 in Appendix A.

We note that this deviation already appears in case of unchanged risk level, and is attributable to the methodological definition of $GCA_t$. By deducting some provision amount in bucket 1 (ie the 12-month expected loss), a structurally optimistic evaluation by NCA$_t$ in case of a delayed risk profile is avoided and a certain conservatism already at loan origination is explicitly introduced.

A comparable discussion of the economic value compared to accounting value in this situation can already be found in Benston and Wall (2005). Although the authors refer to US GAAP generally, their observations are equally valid for current accounting regimes: GAAP values ignore anticipated changes in PD. Varying hazard rates (non-neutral risk profiles) effectively cause distortions of accounting values over time.

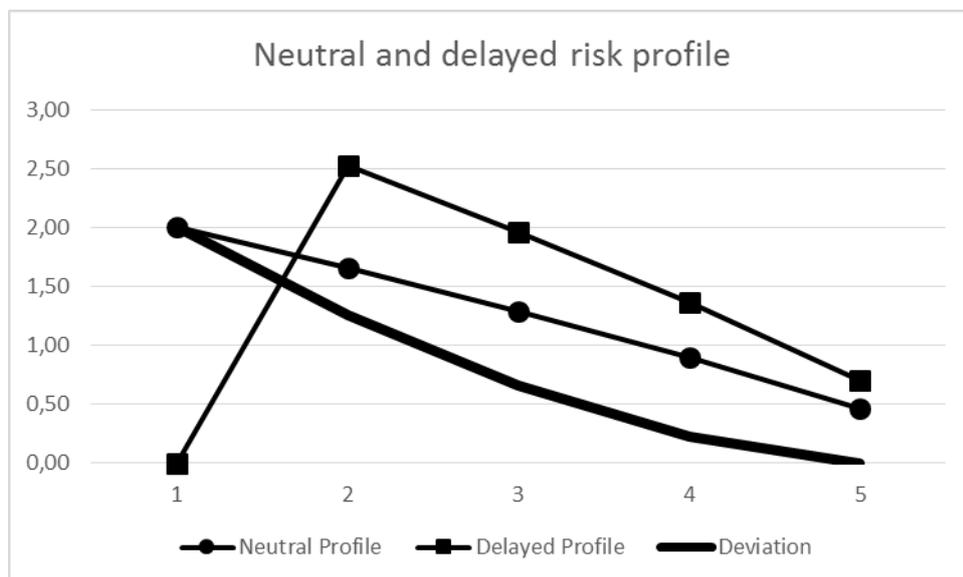

Figure 4.1: Example with different risk profiles and same risk level

The figure compares a neutral and a delayed risk profile for a 5-year closed end loan. The y-axis actually shows the scaled values $p_t r \frac{GCA_t}{GCA_0} = \frac{R_t}{GCA_0}$. The neutral risk profile decreases proportionally with GCA$_t$. For the delayed risk profile zero risk in year one is assumed, offset by higher risk in years 2 to 5 to end up at the same overall risk level. The difference between GCA$_t$ and iACV$_t$ (as % of GCA$_0$) in this case is highest at the end of the first year with a value equivalent to the 12-month expected loss.

Based on a given risk profile the static calibration test is derived. The risk profile itself can be estimated by means of a static pool analysis (vintage analysis) for a homogenous portfolio with similar risk characteristics. Current portfolio trends and other forward-looking information (portfolio quality changes and economic outlooks) may be incorporated. The conservatism (calibration quality) of NCA$_t$ in bucket 1 is derived by direct comparison with iACV$_t$:

$\Delta_t := iACV_t^0 - NCA_t^0$, where all values are discounted with rate $i_{ED}$ to loan origination (t=0).

In case of a neutral risk profile and risk level r, $\Delta_t = iACV_t^0 - NCA_t^0 = GCA_t^0 - NCA_t^0 = EL_t^0 = r\, GCA_t^0$ corresponds to the annualized lifetime expected loss. This conservatism is caused by the structure of the final standard under stable risk conditions and neutral risk profile in any case.

In case of a general risk profile, $\Delta_t = iACV_t^0 - NCA_t^0 = (iACV_t^0 - GCA_t^0) + (GCA_t^0 - NCA_t^0) =$
$= EL_t^0 - \sum_0^{t-1}(r - r_T)\, GCA_T^0$ can be derived when applying Corollary A 3. As a consequence, $NCA_t$ remains conservative as long as $\sum_0^{t-1}(r - r_T)\, GCA_T^0 \leq EL_t^0$ (eg with constant hazard rates $r_T = r$).

The following interpretation may prove useful: $\sum_0^{t-1}(r - r_T)\, GCA_T^0$ corresponds to the cumulative accounting mismatch caused by accruing credit-risk related revenues (as part of the effective interest rate) in line with $GCA_t^0$. The corresponding expenses (actual credit-risk), however, are typically incurred in different accounting periods. The final standard tries to handle this accounting mismatch by introducing a baseline provision of $EL_t^0$ starting with day one.

Remark 4.2: Whether the delay in the risk profile is caused by actual (as in bullet loans) or apparent (caused eg by poor internal processes for risk identification) behavior, is irrelevant. Significantly delayed risk recognition may cause non-conservatism of $NCA_t$ in bucket 1.

**5. $NCA_t$ compared to $iACV_t$: the case of variable risk level**

As long as the risk level in relation to the original risk estimate does not significantly increase, the financial asset may remain in bucket 1. Consequently, the 12-month expected loss has to be recognized as provision amount.

In the definition of this 12-month expected loss provision the final standard applies the regulatory definition: "*12-month expected credit losses are the portion of lifetime expected credit losses that represent the expected credit losses that result from default events on a financial instrument that are possible within the 12 months after the reporting date*". The IASB decided not to choose the annualized expected loss, which would be more in line with the general methodological understanding of the final standard. However, the chosen definition is pragmatic and closely related to the regulatory concept of the shortfall.

This specific choice of the initial provision amount is not methodologically justified. It serves primarily as a buffer to offset the non-conservatism introduced by delayed, non-neutral risk profiles (as shown in the previous section) or to cover small increases in the risk level. It shall ensure that the final standard remains conservative in a practicable range of risk levels.

For a quantification of the effect of an increasing risk level, the modified duration ($D_{mod}$), derived from the Macaulay duration ($D_{Mac}$) of a financial instrument need to be introduced:

$$D_{Mac} := \frac{\sum t\, CF_t/(1+i)^t}{\sum CF_t/(1+i)^t} = \sum \frac{t\, CF_t/(1+i)^t}{GCA_0} = \frac{GCA_0 + GCA_1^0 + \cdots + GCA_T^0 + \cdots}{GCA_0} \text{ and}$$

$$D_{mod} := \frac{D_{Mac}}{1+i} = \frac{GCA_0 + GCA_1^0 + \cdots + GCA_T^0 + \cdots}{(1+i)\, GCA_0} \text{ where } GCA_T^0 = \frac{GCA_T}{(1+i)^T}.$$

In case of the exposure value $iACV_t$ and the respective discount rate $i_{ED}$,

$$D_{mod}^{iACV} = \frac{iACV_0 + iACV_1^0 + \cdots + iACV_T^0 + \cdots}{(1+i_{ED})\, iACV_0}.$$

The modified duration can be interpreted as semi-elasticity (eg Schwaiger, 2014), relating the relative change in present value $GCA_0$ or $iACV_0$ to the absolute change in the applied discount rate $i$ or $i_{ED}$. An increasing risk level (with absolute change $\Delta r$), which has the same impact on the present value as a corresponding increase of the discount rate, leads to a (relative) decrease of $iACV_0$ by $\left(-\Delta r\, D_{mod}^{iACV}\right)$. This modified duration may play an important role in the context of lifetime EAD estimation, which is out of scope of this paper.

On basis of this relationship, the static calibration test from last section to variable risk levels can be generalized for general risk profiles and changing risk level as follows:

$$\Delta_t = iACV_t^0(\text{risk level, risk profile}) - NCA_t^0 =$$
$$= \left(iACV_t^0(\text{risk level, risk profile}) - iACV_t^0(\text{original level, risk profile})\right) +$$
$$+ \left(iACV_t^0(\text{original level, risk profile}) - NCA_t^0\right) =$$
$$= \left(-\Delta r\, D_{mod,t}^{iACV}\, GCA_t^0\right) + \left(EL_t^0 - \sum_0^{t-1}(r - r_T)\, GCA_T^0\right)$$

where $D_{mod,t}^{iACV}$ is the modified duration measured at time t.

Consequently, $\Delta_t$ remains conservative ($\geq 0$) if and only if $\Delta r \leq \frac{EL_t^0 - \sum_0^{t-1}(r-r_T)\, GCA_T^0}{D_{mod,t}^{iACV}\, GCA_t^0}$.

If little conservatism remains after adjusting for the temporal distribution of risk (ie if $\left(EL_t^0 - \sum_0^{t-1}(r - r_T)\, GCA_T^0\right)$ is small) or in case of a large modified duration $D_{mod,t}^{iACV}$ of the financial asset (eg in case of long term contracts), NCA may become non-conservative.

In bucket 2, after detection of a significant risk increase, the quantification of the implicit conservatism is straightforward: the originally expected lifetime losses (based on neutral risk profile) are implicitly reflected in the GCA of the final standard. The requirement of provisions covering actual lifetime expected losses consequently causes the double counting of the originally considered (remaining) lifetime losses in NCA.

Figure 5.1 shows the relationship between $NCA_t$ and $iACV_t$ in a situation with continuously increasing risk level for a non-amortizing loan, term 5 years, and constant gross carrying amount of 100. The y-axis denotes the net value after allowances, whereas time in years is depicted on the x-axis. Expected credit losses increase gradually with time, causing transition to bucket two after 1.5 years in this example. The smooth curve describes the development of the benchmark $iACV^©$. $NCA_t$ (= IFRS 9 in the chart) starts more conservative (caused by the initial 12-months ECL allowance), but gets non-conservative with gradually increasing risk level. Transition to bucket two causes $NCA_t$ to become conservative again, at a much more significant level than in bucket one.

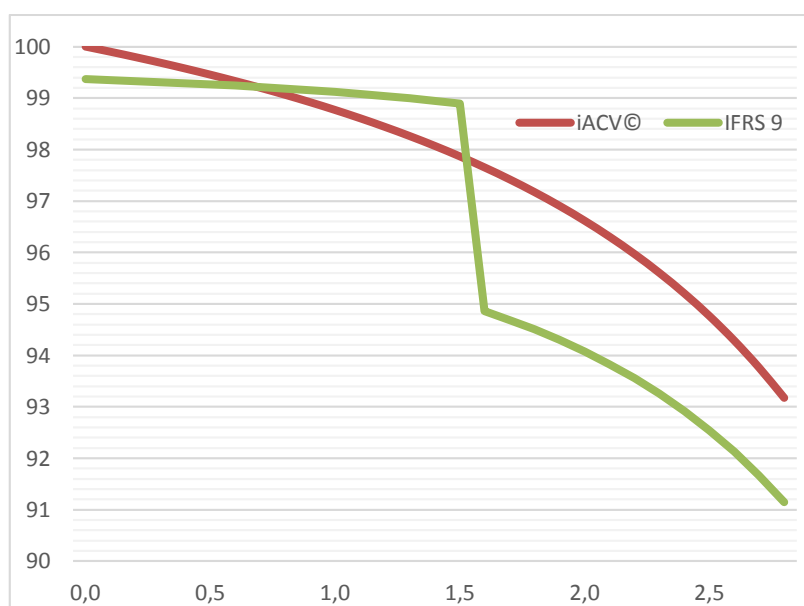

Figure 5.1: Relationship between $NCA_t$ and $iACV_t$ with increasing risk level

The main process responsible for the actual size of conservatism is the timing of transition to bucket 2. If the applied thresholds are overly sensitive to small or temporary changes, and the bucket change is triggered very early, the conservatism in bucket 2 could reach significant amounts. If risk increases are detected too late, however, the ongoing assessment in bucket 1 may become non-conservative as seen in the beginning of this section. An overly sensitive implementation of this transition process may – in the worst case – even question the adequacy of the final standard in the sense of a "best possible representation of the financial asset".

The following example shall make this last statement more intuitive: if a relative increase in risk level by 25% (a PD change from 2% to 2.5%) is already triggering the transition to bucket 2, the relationship between conservatism and provision would be 1:1.25. Ie 80% of the booked provision can be considered potential hidden reserves. Even if the threshold for risk increase is set to a very high level of 500% (PD from 2% to 10%), potential hidden reserves would still amount to 20% of the booked provisions. The correct timing of the transition to bucket 2 is critical to avoid an overly pessimistic creation of potential hidden reserves in bucket 2.

For the practical implementation of this threshold, the final standard gives little practical advice: "*At each reporting date, an entity shall assess whether the credit risk on a financial instrument has increased significantly since initial recognition. When making the assessment, an entity shall use the change in the risk of a default occurring over the expected life of the financial instrument instead of the change in the amount of expected credit losses. To make that assessment, an entity shall compare the risk of a default occurring on the financial instrument as at the reporting date with the risk of a default occurring on the financial instrument as at the date of initial recognition and consider reasonable and supportable information, that is available without undue cost or effort, that is indicative of significant increases in credit risk since initial recognition*".

This definition is methodologically consistent with the final standard by effectively relating the change to bucket 2 to changes in risk level and profile (lifetime loss expectations). The required significance of the change (ie the size if the threshold) and the adequacy of the resulting potential hidden reserves are, however, are not addressed.

Alternatively, the IASB refers to the 1-year regulatory PD in triggering the transition: "*However, the IASB observed that changes in the risk of a default occurring within the next 12 months generally should be a reasonable approximation of changes in the risk of a default occurring over the remaining life of a financial instrument and thus would not be inconsistent with the requirements*". This note may become particularly counterproductive in case of behavior scorings with a high level of month-to-month variability, where it could have a significant impact on provisions in bucket 2.

The rebuttable presumption, taking the 30-day past due event as evidence of "significant increase in credit risk", may have a similar impact on provisions in bucket 2. Alternative transient early warning indicators, with fast response patterns, should be considered with caution for this very reason. Clear focus needs to be on the (statistical) significance of any trigger – pure volatility needs to be avoided while relevant and predictive indicators have to be taken into account as early as possible.

Summarizing, a quantification of the implicit conservatism (potential hidden reserves) introduced by the final standard in buckets 1 and 2 was proposed. An ongoing monitoring of this deviation from the economic value iACV$_t$ is required for a successful implementation of IFRS 9. It highlights overly optimistic or pessimistic evaluations and should improve or ensure comparability of financial assets between entities. Increasing earnings volatility, especially if left unexplained, could otherwise have unintended negative consequences for the financial sector and may even increase structural risks.

**6. Forward looking provisions: The use of forecasts in expected loss estimation**

One of the major differences between regulations and accounting standards is the objective of loss estimates. Regulators generally demand loss estimates to match long-term default rates and expect application of conservative estimates in case of uncertainty. Specifically we find minimum time series requirements and reference to a "through-the-cycle" approach, which is methodologically close to an unconditional loss expectation. Conservatism is included by definition of minimum thresholds, eligibility criteria for collateral or it is explicitly requested through add-ons in case of unreliable or unrepresentative data. This understanding was re-enforced in EBA (2014b), article 49.

The focus of accounting standard setters on the other hand is an economically unbiased "best" estimate and a fair and objective exposure value on basis of (reasonably) available information at the time (forward looking, "point-in-time", conditional loss estimates).

Regulators aim for a certain stability of the minimum capital requirements, while requesting conservatism in case of uncertainty. The general tendency of rating systems to behave cyclical is considered with caution, as is the case with RWA variability (eg EBA 2013). In accounting standards, however, the current best prediction (with reasonable effort) is to be used as estimated asset value. Changing market expectations have to be reflected, ideally in terms of fair value accounting. Even when measuring at amortized cost through forward looking loss estimates, the current outlook has to be incorporated by consideration of internal and external factors.

As internal factors we understand changes to risk-related processes, under the responsibility and influence of the management, which may have a sustainable and significant effect on the overall risk level or on the temporal distribution of losses. These impacts can be considered in form of an adjustment to the loss estimates. Relevant process measures must be defined and monitored to reconfirm the validity of these adjustments on an ongoing basis. There may be good arguments for risk improvements (such as improved credit rating systems, higher collateral requirements, strengthened underwriting to reduce high-risk customer segments, etc.), but also potential risk increases after long periods of low losses have to be considered. As example, significant above-average growth in an otherwise stable segment might indicate potential yet hidden risk increases.

External factors are found predominantly in the area of economic macro developments and tend to be cyclical and have to be considered when deriving loss estimates for IFRS 9. The challenge will be the separation of overly optimistic or pessimistic market sentiments (noise) from the actually valid and backtestable content of predictions (trends). Historic evidence proves that the prediction (and the acceptance of these predictions by the industry) of the timing and depth of an upcoming crisis situation as needed to estimate the (lifetime) expected loss is extremely difficult.

An interesting analysis on basis of the 2008/2009 crisis can be found in Ivantsov (2013). He demonstrated that – few weeks before the default of Lehman Brothers – the depth of the subsequent economic impact was seriously underestimated. Although an upcoming downturn was already expected, its systemic impact, however, was ultimately not recognized. As a matter of fact, economic forecasts seem to be only reliable with respect to relatively short-term trend indications (a few quarters), but fail seriously with respect to forecasting the depth of upcoming extreme events.

This uncertainty has to be considered when implementing IFRS 9 impairment standards. An uncritical adoption of long-term forecasts would almost certainly cause excessive cyclicality and thus may well increase the likelihood of credit bubbles and subsequent losses in case the economy finally turns. Thus, the volatility caused by the predictive aspect in risk estimates needs to be carefully controlled.

For easier understanding of this effect let us take the following naïve example: In a typical portfolio with annual risk cost of 100 million euros (long-term average, regulatory point of view), only 50 million euro were realized annually in the past two years due to positive economic developments. This may easily be in line with a general economic capital model, where mode and median (to be considered typical annual results) are significantly below the mean of the loss distribution (corresponding to the expected loss). The general economic forecasts continue to expect positive development. However, we know on basis of not too long history that realized risk costs amounted to around 500 million euro in recent crisis situations. The loss estimate (which according to IFRS 9 has to be based on probability-weighted outcomes) must take into account that such an event will realize with a probability of 5-20%. Thus the fair estimate, even after taking into account the actual positive economic outlook, is closer to 100 million euros (= 50 million * 90% + 500 million * 10%) and may be reasonably close to the long-term regulatory estimate.

Statistical verification of these predictions in credit risk through backtesting poses the next analytical challenge: risk costs generally have to serve a variety of managerial objectives, as already mentioned in Section 2, and do not have the statistical properties needed for proper testing procedures.

To objectively test the validity of economic forecasts, undistorted and frequently (quarterly or monthly) available estimates of realized credit risk are necessary. One measurement to satisfy these requirements was derived in Reitgruber (2013), through Impact of Risk$^©$ (IoR$^©$). A short introduction is included in Appendix B. The application of this framework allows for a high frequent (monthly) measurement of credit risk. In contrast to the classical credit risk measurement via direct write-offs or impairment changes, the timing of risk realization is largely defined by the regulatory definition of default and is robust with respect to manual processes and management input.

To verify the effect of external (or internal) factors on credit risk, the PL Dashboard from the IoR© framework will be primarily used. In this measure, realized credit risk is identified through the regulatory definition of default and quantified by EAD and LGD. The regulatory definition of default is largely standardized by definition of specific risk signals and the 90-day (or 180-day in case) past due threshold. Consequently, we can assume a fairly homogenous application across different portfolios and even across financial institutions. Another recent consultation paper by EBA (2014a) is attempting to further align and strengthen the default definition – although this is most likely causing inconsistencies in the default time series right after implementation, it will ultimately support an even more objective application of Impact of Risk.

Remark 6.1: For the application of the PL Dashboard the accuracy of the EAD and LGD estimates is less critical than the default identification process: the product EAD x LGD is practically applied only as a weighting function in a classical backtest of default rates. The validation of the LGD model itself is performed through a separate NPL Dashboard and the EAD model is validated through the annual version of the PL Dashboard as outlined below with $\Delta_{EAD}$.

In case of an annual observation period the PL Dashboard can be described as

$$\text{PL Dashboard} = \text{EL}_{\text{new NPL}}^{\text{EOP}} + \text{wo}_{\text{new NPL}} - \text{EL}_{\text{PL}}^{\text{BOP}} \text{ or}$$

$$\text{EL}_{\text{new NPL}}^{\text{EOP}} = \text{EL}_{\text{PL}}^{\text{BOP}} - \text{wo}_{\text{new NPL}} + \text{PL Dashboard}.$$

To separate impacts from PD, EAD and LGD the split version is derived as follows:

$$\text{PL Dashboard} = \sum_{\text{new NPL}} \left(\text{EAD}^{\text{BOP}} \text{LGD}_{\text{PL}}^{\text{BOP}}\right) - \text{EL}_{\text{PL}}^{\text{BOP}} +$$

$$+ \sum_{\text{new NPL}} \left(\text{EAD}^{\text{DEF}} \text{LGD}_{\text{PL}}^{\text{DEF}} - \text{EAD}^{\text{BOP}} \text{LGD}_{\text{PL}}^{\text{BOP}}\right) +$$

$$+ \sum_{\text{new NPL}} \left(\text{EAD}^{\text{EOP}} \text{LGD}_{\text{NPL}}^{\text{EOP}} - \text{EAD}^{\text{DEF}} \text{LGD}_{\text{PL}}^{\text{DEF}}\right)$$

BOP and EOP represent the beginning and end of the observation period and DEF denotes the time of default. Non-performing loans are differentiated whether they were non-performing at the beginning of the observation period ("old NPL") or not ("new NPL").

The first term ($\Delta_{\text{PD}} := \sum_{\text{new Defaults}} \left(\text{EAD}^{\text{BOP}} \text{LGD}_{\text{PL}}^{\text{BOP}}\right) - \text{EL}_{\text{PL}}^{\text{BOP}}$) corresponds to the traditional impact based PD backtest with weights $\text{EAD}^{\text{BOP}} \text{LGD}_{\text{PL}}^{\text{BOP}}$.

The middle term ($\Delta_{\text{EAD}} := \sum_{\text{new Defaults}} \left(\text{EAD}^{\text{DEF}} \text{LGD}_{\text{PL}}^{\text{DEF}} - \text{EAD}^{\text{BOP}} \text{LGD}_{\text{PL}}^{\text{BOP}}\right)$) describes an EAD backtest with weights $\text{LGD}_{\text{PL}}^{\text{BOP}}$, corrected for collateral repossessions and other LGD changes short before the default event. $LGD_{PL}^{DEF}$ has to be in line with the LGD method for performing loans, in case there is a methodological discontinuity to the LGD method for non-performing loans.

The last term ($\Delta_{LGD} := \sum_{\text{new Defaults}} (EAD^{EOP} LGD_{NPL}^{EOP} - EAD^{DEF} LGD_{PL}^{DEF})$) compares LGD immediately before ($LGD_{PL}^{DEF}$) and after ($LGD_{NPL}^{EOP}$) default and should generally be a small correction or error term. A large deviation may indicate discontinuity of the LGD method right around and after default, or may even highlight an incorrect estimate of the early cure rate in LGD.

Measurements of credit risk are mostly based on annual default rates. In case of a sufficiently stable portfolio, monthly or quarterly PDs can be derived by down-scaling annual values. Under this assumption, the PL Dashboard can be described as follows:

$$\text{PL Dashboard}_M = EL_{\text{new NPL}}^{EOP} + wo_{\text{new NPL}} - EL_{PL}^{BOP}/12 \text{ or}$$

$$EL_{\text{new NPL}}^{EOP} = EL_{PL}^{BOP}/12 - wo_{\text{new NPL}} + \text{PL Dashboard}_M.$$

Main advantage of a shorter observation period is a significant reduction of the deviation caused by EAD or LGD in $\Delta_{EAD}$ und $\Delta_{LGD}$. Therefore the monthly or quarterly versions of the PL Dashboard may well be sufficient for further analysis of the cyclical effects, instead of the more exact $\Delta_{PD}$. Should this approximation prove to be inadequate for the analyzed portfolio, the more detailed split representation as described above has to be applied.

Additionally, the influence of $EL_{PL}^{BOP}$ in a stable portfolio – in comparison to the generally high stochastic component in actual credit risk – is typically low. For a more detailed analysis of the dependency structures the monthly or quarterly time series of losses $\text{Loss} := EL_{\text{new NPL}}^{EOP} + wo_{\text{new NPL}}$ should be sufficient.

The Loss time series can be interpreted as the rhythm or pulse of credit risk in a medium to high default portfolio. An understanding of its stochastic properties (distribution, dependency structure) will provide another valuable benchmark for EL estimates. With this measure, currently observed RWA differences for seemingly comparable portfolios (eg EBA (2013) and further upcoming analyses by regulators and internal analyses by PECDC (2013a) or PECDC (2013b)) might find its justification by different Loss distributions.

On basis of Loss a test on the loss distribution, and in particular the significance of correlation effects, can be developed. The null hypothesis of the naïve portfolio model with independent binomially distributed defaults provides a check of the plausibility of the volatility of actual credit risk. This hypothesis represents a lower limit of stochastic effects in credit risk and helps to separate trends from purely random statistically significant month results. This property makes it highly relevant for budgeting processes and actual vs plan performance comparisons.

During calm economic periods the correlation of defaults (asset correlation) is expected to be low. Consequently, the null hypothesis of the naïve portfolio model can be used to test for of asset correlations. With monthly data points available covering a period of 5 to 10 years, a common crisis scenario (in the range of the 80% -95% quantile) can be effectively tested. The suggested framework provides a starting point for further empirical analyses and academic research based on actual portfolio developments and may be used to justify (or reject) cycle adjustments for IFRS 9 impairment accounting.

**7. LGD estimation in bucket 3 and an alternative backtesting approach**

For defaulted exposures in bucket 3 there are obvious methodological differences between the final accounting standard and the regulatory requirements on LGD estimation. Appendix A of the final standard defines the amount of credit risk as follows: "*The difference between all contractual cash flows that are due to an entity in accordance with the contract and all the cash flows that the entity expects to receive (ie all cash shortfalls), discounted at the original effective interest rate*".

Regulatory LGD estimates follow a different approach: the applicable discount rate may vary in a large range from risk free rates to effective interest rates. High attention is given to the cost structure after default, conservatism assumptions and corrections for downturn situations ("Downturn LGD").

A benchmark analysis performed by EBA (2013) confirms the large bandwidth of actually applied discount rates: "*The discount rates vary from below 2.5% to 12.5% due to difference in risk-free rate (different time horizon taken into account) and additional risk premium*" in SME segments and mortgage portfolios. A similar result can be found in PECDC (2013a). The academic discussion on appropriateness of certain discount rates to estimate economic LGDs also has already significant history. A good overview can be found in Maclachlan (2005).

Due to the generally long duration of a recovery process, the choice of the discount rate is definitely one key driver for the LGD estimate and may cause significant RWA differences. A clear definition such as the choice of the effective interest rate in the final standard should lead to more comparable results than the wide range allowed in the regulatory context.

When comparing amortized cost accounting with an idealized fair value as done in Section 3, the changing cost and capital structure after default may cause inconsistencies. Defaulted transactions may require more intensive collections efforts, justifying increased cost allocation. On the other hand, significantly reduced capital requirements – while assuming unchanged profit expectations by unit of capital – may offset this effect. The total of these differences may cause potential hidden reserves or hidden liabilities in bucket 3.

The conservatism considerations from Section 5, where we have shown that the original risk level is already included in the gross exposure value (GCA), are less relevant in bucket 3. Although bucket 3 contains the same type of implicit conservatism, compared to the total loss expectation for already defaulted transactions, this structural conservatism in the final standard is comparably small.

The final accounting standard summarizes the following arguments for the choice of the discount rate in its basis for conclusions: *"(a) the effective interest rate is the conceptually correct rate and is consistent with amortized cost measurement; (b) it limits the range of rates that an entity can use when discounting cash shortfalls, thereby limiting the potential for manipulation; (c) it enhances comparability between entities; and (d) it avoids the adjustment that arises when financial assets become credit-impaired (interest revenue is required to be calculated on the carrying amount net of expected credit losses) if a rate other than the effective interest rate has been used up to that point"*.

When applying the IoR© framework, the calibration of the LGD is observable in the NPL Dashboard:

$$\text{NPL Dashboard} := \text{EL}^{\text{EOP}}_{\text{old NPL}} + \text{wo}_{\text{old NPL}} - \text{EL}^{\text{BOP}}_{\text{NPL}} \text{ or}$$

$$\text{EL}^{\text{EOP}}_{\text{old NPL}} = \text{EL}^{\text{BOP}}_{\text{NPL}} - \text{wo}_{\text{old NPL}} + \text{NPL Dashboard}.$$

As shown in Reitgruber (2013), the NPL Dashboard under regulatory requirements and unbiased LGD estimation (ie correct prediction of cash flows) is non-zero. Discounting and indirect cost components are distorting factors. On basis of IFRS 9 accounting standard, this distortion amounts to the (negative) unwinding interest in the respective accounting period. Only after correcting the NPL Dashboard for unwinding interest, or by accruing unwinding interest in GCA, the NPL Dashboard becomes unbiased. Proves for these statements are given in Lemmata B 1 and B 2 in Appendix B.

Remark 7.1: The specific role of the 12-month observation period in the PL Dashboard (caused by the PD definition) is irrelevant for the NPL Dashboard. Consequently, the NPL Dashboard can be applied without adjustments to any observation period.

The application of the NPL Dashboard is possible in two ways. The methodological perspective will focus on the vintage (static pool) analysis, where the cumulative LGD is backtested against an extended time series. The monthly (moving window) representation will be more convenient for portfolio monitoring, where it supports monthly comparisons of actual results with targets.

Before we go into the detailed application of the NPL Dashboard, we need to split the net exposure value NCA of defaulted transactions into its components:

$NCA_{Coll}$ := the material value of a collateral after haircut, corresponding to the expected discounted net cash flow from collateral. If the material value is not part of the LGD method (eg in case of an overall LGD), this term vanishes.

$NCA_{Unsec}$ := $(1 - LGD_u)$ $(GCA - NCA_{Coll})$ = expected discounted cash flows based on the unsecured exposure. $LGD_u$ denotes LGD in percent of the unsecured portion of a transaction (the unsecured LGD). In case of an overall LGD method, $LGD_u$ = LGD and $NCA_{Coll}$ = 0. $NCA_{Unsec}$ does not consider possible PD reductions through guarantees or other cases with PD < 100% caused by cured transactions, as long as they continue to be reported as defaulted in bucket 3.

$NCA_{G'tee}$ := $(1-PD)$ $LGD_u$ $(GCA - NCA_{Coll})$, in case of PD < 100%. This amount quantifies expected cash flows caused by guarantor or cures beyond $NCA_{Unsec}$ and $NCA_{Coll}$, as long as these transactions continue to be reported as defaulted in bucket 3. The PD in this case corresponds to the default rate of the guarantor or the current default rate of the cured transaction.

For cured transactions, reported as non-defaulted in bucket 1 or 2, NCA is not further split into its components and shall be reported as $NCA_{Cure}$. If curing happens frequently, a further split into collateral and unsecured might be appropriate. Neglecting cured accounts could distort the resulting vintage analysis in case of iterative defaults and should be avoided.

Remark 7.2: The scaling factor $LGD_u$ highlights that the focus of $NCA_{G'tee}$ are incremental cash flows covered by the guarantee only. A high $LGD_u$ close to 100% means that a large portion of NCA is to be covered by the presence of the guarantor. A low $LGD_u$ means that most of the expected cash flow is borne by the primary client already.

We prove the relationship NCA = $NCA_{Coll}$ + $NCA_{Unsec}$ + $NCA_{G'tee}$ for defaulted portfolios in Lemma B 3 in Appendix B. The separation of NCA into expected cash flows based on collateral, guarantors and cures, and other (unsecured), allows for a detailed and transparent analysis of defaulted loans and the underlying collections processes.

The application of the NPL Dashboard in the framework of a vintage analysis (static pool analysis) is performed as follows: a portfolio of transactions, which defaulted in a certain observation period (eg 1.1.-31.12.2008 as "static pool") is kept constant. At the first future observation date (eg 31.12.2008 as time "0") the GCA is separated into $GCA^0$, $NCA_{Coll}^0$, $NCA_{Unsec}^0$, $NCA_{G'tee}^0$ and the residual $EL^0$.

The NPL Dashboard applied at BOP=t-1 and EOP=t leads to the following relationship:

$EL_{old\ NPL}^t = EL_{NPL}^{t-1} - wo_{old\ NPL}^t + NPL\ Dashboard^t$, or

$EL^t = EL^{t-1} - wo^t + NPL\ Dashboard^t$, since there are no new NPLs in the static pool by definition.

The cumulative version useful for the vintage analysis is easily derived:

$$EL^T = EL^0 - \sum_1^T wo^t + \sum_1^T NPL\ Dashboard^t\ or$$

$$TEL^T := EL^T + \sum_1^T wo^t + \sum_1^T i\ NCA^t = EL^0 + \sum_1^T (NPL\ Dashboard^t + i\ NCA^t).$$

TEL denotes the total expected loss including already incurred write-offs and corrected for unwinding interest on the static pool. Based on the time series $TEL^T$, combined with a stochastic model for the cumulative random process $\sum_1^T (NPL\ Dashboard^t + i\ NCA^t)$ with mean zero (proven in Lemma B 1 in Appendix B), a calibration test for the LGD method can be derived. Analysis of the time series $NCA_{Coll}^T$, $NCA_{Unsec}^T$ and $NCA_{G'tee}^T$ allow for a quantification of the efficiency and average duration of internal collections processes for secured or unsecured cash flows. Deviations can be allocated to specific periods and support a quantification of the impact of cyclical situations or internal process changes on the LGD.

This framework allows for improved downturn LGD analyses, as it clearly separates actual cash flows, collections processes and default periods. An interesting analysis on workout-based downturn LGD was published by PECDC (2013c), already indicating that the timing of cash flows might be more critical than the timing of defaults for a clean identification of the downturn LGD effect. Further empirical analyses and academic research will be necessary to fully understand the stochastic properties of these time series.

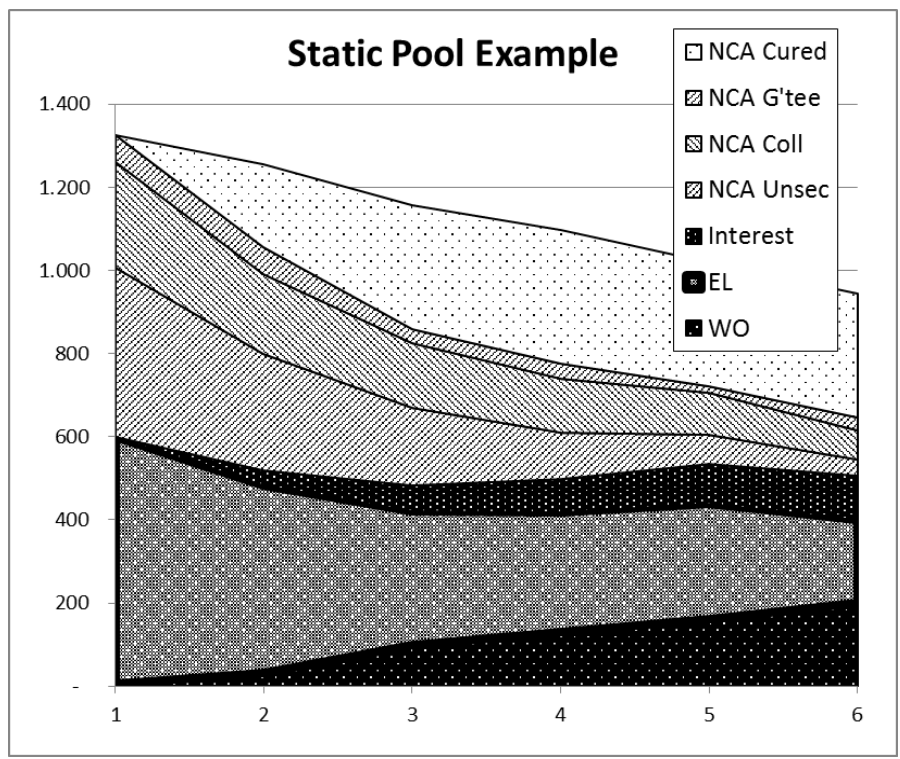

Figure 7.1: Example of a 6-year static pool analysis (symbolic but typical data)

In this example, the speed of reducing $NCA_{Unsec}$ is clearly visible. Early cures significantly contribute to this effect in the first 2-3 years (increasing $NCA_{Cured}$). Collaterals instead seem to stay on book much longer – $NCA_{Coll}$ remains on relatively high level for longer periods. The effect of guarantors $NCA_{G'tee}$ however seems to have little material impact for being relevant for LGD. A sudden and strong increase however could by a sign for changes in the portfolio structure or collections practices. Hackl (2014) developed a first empirical study in line with this approach.

The unwinding interest portion of NCA ("interest") was depicted separately. A stable (flat, horizontal) TEL as sum of WO, EL and interest (lower 3 areas) is an indication of unbiased LGD estimates.

The alternative application of the NPL Dashboard in the monthly moving average context can be used to analyse credit risk movements during short observation periods (typically monthly or quarterly, or annual as well) in an ongoing portfolio monitoring context. Monitoring the indicators NPL Dashboard$^t$ (not cumulated), the stocks $NCA_{Coll}^t$, $NCA_{Unsec}^t$ and $NCA_{G'tee}^t$ (not restricted to certain static pools) and the movements of $NCA_{Coll}^t$, $NCA_{Unsec}^t$ and $NCA_{G'tee}^t$ (corrected for new defaults, comparable to the NPL Dashboard) will clearly highlight short-termed changes of risk levels and their root causes.

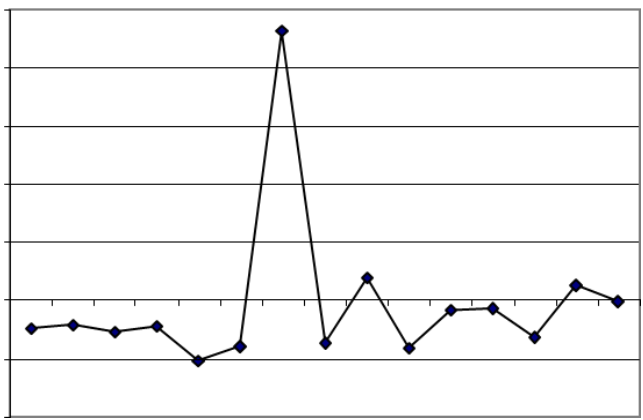

Figure 7.2: Symbolic example of the monthly NPL Dashboard with LGD method adjustment (peak)

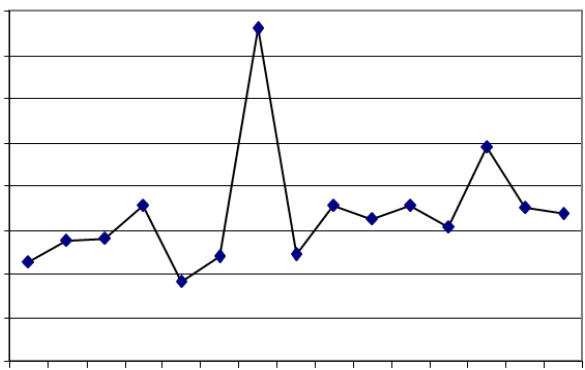

Figure 7.3: Symbolic example of monthly NCA movements with LGD method adjustment (peak)

Ongoing positive or negative deviations in the NPL Dashboard may indicate aggressive or conservative LGD estimates. Trends or structural breaks in $NCA_{Coll}^t$, $NCA_{Unsec}^t$ and $NCA_{G'tee}^t$ highlight possible changes in the portfolio distribution or of collections processes. In this sense, the suggested key risk indicators should be included in any extensive risk reporting.

## 8. Summary

Although it is certainly too early for a general statement about the methodological consequences of the finally released IFRS 9 accounting standards for impairment, discussions on the following three topics were presented in this paper:

First topic was the discussion of potential hidden reserves (conservatism) or liabilities in the upcoming IFRS 9 impairment standard in buckets 1 and 2 due to the chosen compromises. The methodological understanding of these choices and the quantification of the resulting deviations from the economic value will be necessary to separate the effects of implementation details (such as the transition process from bucket 1 to bucket 2) from actual risk performance.

In the next section, the framework Impact of Risk$^©$ is applied to effectively quantify and validate economic impacts on credit risk. Further empirical analyzes must follow before a definitive statement of significance and validity for specific portfolios is possible. In any case, a deeper analysis of portfolio model behavior in the lower 80% - 95% quantiles of the loss distribution will be necessary to justify forward looking assumptions in impairment accounting. Otherwise the new accounting standards could introduce significant P&L volatility, which in turn may even cause additional systemic risk.

The closing section finally focuses on the NPL Dashboard, which is applied to validate LGD methods in a static pool as well as a moving window context. Its systematic application – in the framework of portfolio monitoring and in form of a static pool analysis for more structural studies – will make it possible to quantify the influence of internal and external factors on the LGD.

In this sense we are methodically just at the beginning of a journey that may well provide a lot of insight during the coming few years.

**Appendix A: The relationship between default events and the quantification of (actual) credit risk**

The move from buckets 1 or 2 (performing portfolio) to bucket 3 (non-performing) is the key threshold in credit risk accounting in terms of balance sheet representation, income recognition and in the stochastic structure of credit risk under IFRS 9 as well as under IAS 39:

| Bucket | 1 – low risk level or small increase only | 2 – significantly increased loss | 3 – realized loss event |
|---|---|---|---|
| Provisioning | 12 month expected loss | Lifetime expected loss | Lifetime expected loss |
| Interest accrual | Effective interest rate based on GCA (gross) | Effective interest rate based on GCA (gross) | Effective interest based on amortized cost (NCA, net) |

Table A.1: Bucket overview in the final standard

In bucket 1 and 2 we assume contractually compliant repayment of the obligation. Interest gets charged to the transaction and respective income is recognized on the revenue side. In most cases, the PD will be a low single digit figure. Stochastic properties of credit risk can be characterized as skewed: limited positive deviations, with a fat tail on negative side. Portfolio models are applied to simulate these extreme risk events through dependency structures (eg asset correlations). Summarizing, credit risk is mostly characterized by rare, but significant, events, with the potential to wipe out many years of income or even the bare existence of the institution. Consequently, the unexpected loss component is crucial for the capital requirement in practice in bucket 1 or 2.

In bucket 3 however, credit risk is characterized by increased influence of the expected loss. The distribution is less skew, positive deviations are possible as well as negative. The bimodal properties visible in LGD for non-performing loans are stochastically much less disturbing than the dependency structures and likelihood of extreme events in bucket 1 or 2.

To fully backtest credit risk one would have to wait until the whole portfolio is either paid back or written-off. In order to avoid this situation in practical application and to remove the dependency on this lengthy procedure, the observation of (actual) credit risk can be divided into two phases:

In buckets 1 and 2 we define as the time of realization of credit risk the transition to bucket 3 = non-performing. The amount of realized credit risk is determined by the estimates EAD x LGD (= expected lifetime loss, conditional on the default event) at default in this case.

In bucket 3, the expected lifetime loss is not recognized as receivable anymore and interest accrual will be performed only to the net amount (unwinding). In this bucket, the comparison of initial estimates EAD x LGD with the actual (discounted) cash flows is performed, verifying the adequacy of the lifetime loss estimate used in assessing actual credit risk from the first step. These methodological conventions are in line with standard financial theories (eg Schwaiger 2014).

A proper backtest for the LGD in the NPL portfolio would require long time series, up to the total liquidation of a portfolio. The NPL Dashboard takes an alternative route, corresponding to a piecewise backtest of the LGD slope. Indirect costs and discounting requirements have to be offset through a correction term in the NPL Dashboard, to allow for an unbiased view on the deviations.

In order to describe the methodological relationship between the final standard and the ED 2009, the definitions from Sections 3 to 5 are used.

**Lemma A 1**: In case of a neutral risk profile $GCA_t = iACV_t$ holds for all periods t.

**Proof of Lemma A 1** with mathematical induction:

$GCA_0 = iACV_0$ = the asset value at loan origination (Definition in the final standard and in ED 2009).

If for all T from 0 to t with t ≥ 0 the relationship $GCA_T = iACV_T$ holds true, we prove for t+1:

$$GCA_{t+1} = (1 + i)\, GCA_t - CF_t = (1 + i)\, iACV_t - CF_t =$$

$$= (1 + i - r)\, iACV_t - (CF_t - r\, iACV_t) = iACV_{t+1},$$

as $(i - r)$ is the discount rate applied when calculating $iACV_t$ and $(CF_t - r\, iACV_t)$ corresponds to the expected cash flows (net of expected losses) under ED 2009 and neutral risk profile.        q.e.d.

Define $GCA_t^0$, $CF_t^0$ and $R_t^0$ to be the respective present values discounted with rate $i$ to loan origination in line with the final standard, the absolute risk profile $r_t = \frac{R_t}{GCA_t} = \frac{R_t^0}{GCA_t^0}$ and the relative risk profile $p_t = \frac{r_t}{r}$. The neutral risk profile is finally characterized by $p_t = 100\%$ (ie $r_t = r$) for all t.

**Lemma A 2**: A general risk profile $p_t$ has the same risk level as the neutral risk profile if and only if $\frac{\sum p_t GCA_t^0}{\sum GCA_t^0} = 100\%$ (summation goes from 0 to infinity, ie until final liquidation).

Comment: This condition can be used as norming requirement for general risk profiles.

**Poof of Lemma A 2**: Identical risk level means that the sum of (discounted) expected losses must be identical. Since the effective interest rate depends on contractual cash flows only, the same discount rate is to be applied for both risk profiles. Consequently, $\sum R_t^0 = r \sum GCA_t^0$ holds for the neutral risk profile, and $\sum R_t^0 = \sum r_t GCA_t^0$ holds for the general risk profile. It follows that $r \sum GCA_t^0 = \sum r_t GCA_t^0$, or $\sum GCA_t^0 = \sum p_t GCA_t^0$, or $1 = \frac{\sum p_t GCA_t^0}{\sum GCA_t^0}$.        q.e.d.

The relationship $\sum R_t^0 = r \sum GCA_t^0$ can be further used to derive the impact of a general risk profile at time T on the exposure value. For that purpose, the notion of identical risk levels has to be adjusted slightly to the iACV© context: the discount rate $i_{ED}$ is to be applied instead of $i$

**Corollary A 3:**

$$GCA_T^0 - iACV_T^0 = \sum_0^{T-1}(r - r_t)\, GCA_t^0$$

**Proof of Corollary A 3:**

$$GCA_T^0 - iACV_T^0 = \sum_T^\infty (CF_t^0 - r\, GCA_t^0) - \sum_T^\infty (CF_t^0 - R_t^0) =$$

$$= \sum_T^\infty R_t^0 - r \sum_T^\infty GCA_t^0 = r \sum_0^{T-1} GCA_t^0 - \sum_0^{T-1} R_t^0 = \sum_0^{T-1}(r - r_t)\, GCA_t^0 \qquad \text{q.e.d.}$$

Comment: $GCA_T^0$ is written as $iACV_T^0$ with neutral risk profile and discount rate $i_{ED}$.

This means that the difference between the GCA in the final standard to the economic value as defined in ED 2009 corresponds to the cumulative difference of risk level and absolute risk profile, weighted by $GCA_t^0$. If eg losses are delayed by one year (ie $r_0 = 0\%$) the maximum deviation after one year corresponds to the 12-month expected loss $r\, GCA_0^0$.

**Appendix B: Introduction to Impact of Risk©:**

Impact of Risk© (IoR©) as measure to quantify the credit risk is derived on basis of regulatory concepts in the advanced IRB approach. It measures the influence of credit risk on profit and loss (traditional risk costs) combined with the immediate effect on capital requirements through the shortfall. This framework is described in detail in Reitgruber (2013).

Remark B 1: IoR© is derived on basis of the regulatory IRBA concept, where it has an immediate operational counterpart by representing certain actual changes in capital requirement. Nevertheless it can be considered a highly relevant key risk indicator in non-IRBA credit risk segments as well, as long as models for expected loss are used to monitor and estimate credit risk.

The quantity IoR© is defined as follows:

IoR := Cost of Risk (CoR) + ΔShortfall (SF)

CoR denotes the traditional impact of risk on profit and loss, covering changes to the provisioning account and direct write-offs. The shortfall corresponds to the difference between provisions (credit risk adjustments) and expected loss, as defined in the CRR (BIS 2013b). It leads to incremental tier 1 capital requirement (if negative) or potential tier 2 capital (if positive).

According to Lemma 2.3 in Reitgruber (2013), IoR© relates to the expected loss as follows:

IoR = $EL^{EOP} - EL^{BOP}$ + wo

EL denotes the total expected loss based on the regulatory definition, covering 12-month expected loss on performing and the lifetime expected loss on non-performing portfolio, and wo denotes all write-offs in the observation period (direct writeoffs or through usage of provisions). For application in steering and monitoring, the equivalent relationship may be more adequate:

$$EL^{EOP} = EL^{BOP} + IoR - wo$$

This relationship highlights that the analysis of IoR$^©$ is key to explain monthly or quarterly movements of the expected loss and may serve as a basis for credit risk predictions.

According to Lemma 3.1 in Reitgruber (2013) IoR$^©$ can be split into the following components:

$$IoR = EL_{PL}^{EOP} + \text{PL Dashboard} + \text{NPL Dashboard},$$

with PL Dashboard $:= EL_{new\ NPL}^{EOP} + wo_{new\ NPL} - EL_{PL}^{BOP}$

and NPL Dashboard $:= EL_{old\ NPL}^{EOP} + wo_{old\ NPL} - EL_{NPL}^{BOP}$.

From this decomposition, the static and dynamic views on IoR$^©$ are easily derived:

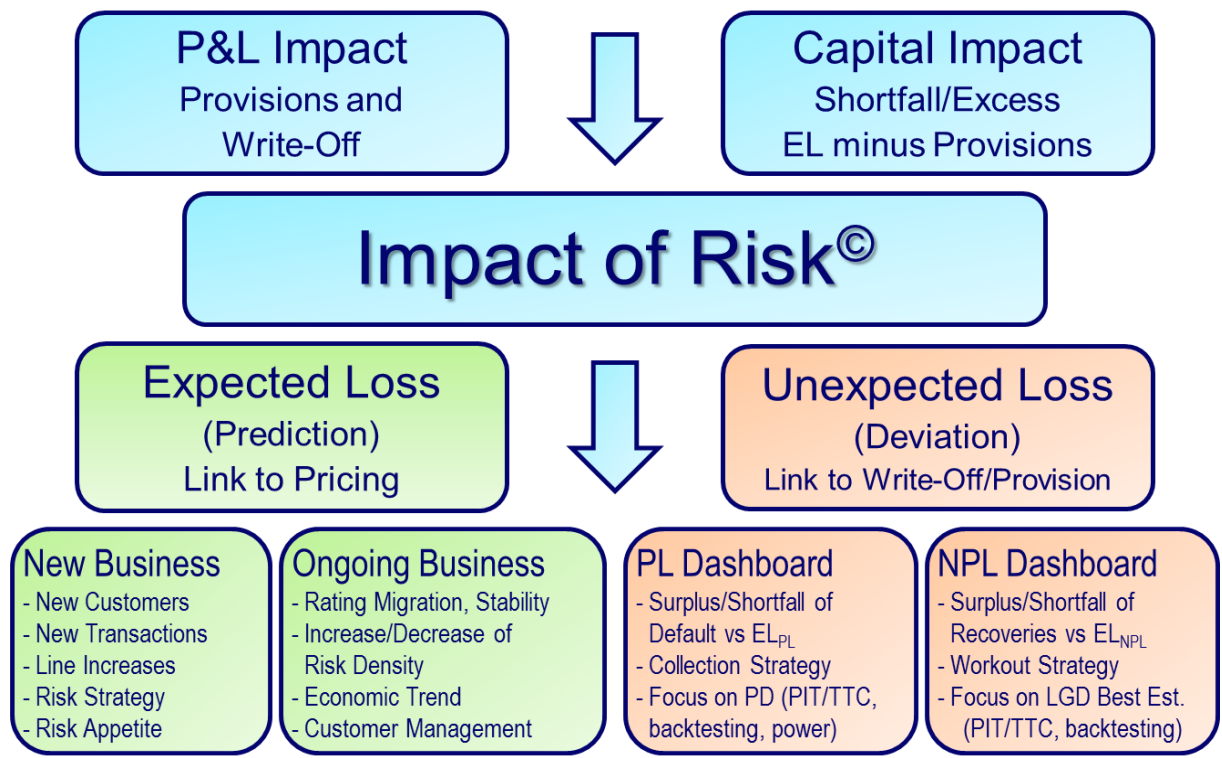

Figure B.1: Static view on IoR$^©$

Figure B.2: Dynamic view on IoR©

The appendix closes with the outstanding proofs from Section 7:

**Lemma B 1**: NPL Dashboard $= -\,i\, \text{NCA}_{\text{NPL}}^{\text{BOP}}$ = the amount of unwinding interest under the assumption of an unbiased cash flow estimate (ie unbiased LGD estimate) for non-performing loans under discounting conditions of the final standard, unwinding of interest applied to NCA and non-accrual of interest applied to GCA.

**Proof of Lemma B 1**:

$$\text{EL}_{\text{NPL}}^{\text{BOP}} = \text{GCA}_{\text{NPL}}^{\text{BOP}} - \text{Rec}_1 \frac{1}{(1+i)} - \text{Rec}_2 \frac{1}{(1+i)^2} - \text{Rec}_3 \frac{1}{(1+i)^3} - \cdots$$

and

$$\text{EL}_{\text{oldNPL}}^{\text{EOP}} = \text{GCA}_{\text{oldNPL}}^{\text{EOP}} - \text{Rec}_2 \frac{1}{(1+i)} - \text{Rec}_3 \frac{1}{(1+i)^2} - \cdots$$

hold, where i is the rate used for discounting in the LGD estimation. $\text{Rec}_t = \text{CF}_t - R_t$ are expected cash flows to be received at the end of year t for the portfolio under consideration, net of direct expenses charged on the account. The NPL Dashboard can be transformed as follows:

$$\text{NPL Dashboard} = \text{EL}_{\text{old NPL}}^{\text{EOP}} + \text{wo}_{\text{old NPL}} - \text{EL}_{\text{NPL}}^{\text{BOP}} = -\,i\, \text{NCA}_{\text{NPL}}^{\text{BOP}}$$

$$= \text{GCA}_{\text{oldNPL}}^{\text{EOP}} - \text{Rec}_2 \frac{1}{(1+i)} - \text{Rec}_3 \frac{1}{(1+i)^2} - \cdots + \text{wo}_{\text{old NPL}} -$$

$$- \text{GCA}_{\text{NPL}}^{\text{BOP}} + \text{Rec}_1 \frac{1}{(1+i)} + \text{Rec}_2 \frac{1}{(1+i)^2} + \text{Rec}_3 \frac{1}{(1+i)^3} + \cdots =$$

$$= \text{GCA}_{\text{oldNPL}}^{\text{EOP}} - \text{GCA}_{\text{NPL}}^{\text{BOP}} + \text{Rec}_1 \frac{1}{(1+i)} + \text{wo}_{\text{old NPL}} - \text{Rec}_2 \frac{1+i-1}{(1+i)^2} - \text{Rec}_3 \frac{1+i-1}{(1+i)^3} - \cdots =$$

$$= \text{GCA}_{\text{oldNPL}}^{\text{EOP}} - \text{GCA}_{\text{NPL}}^{\text{BOP}} + \text{Rec}_1 + \text{wo}_{\text{old NPL}} - i \left( \text{Rec}_1 \frac{1}{(1+i)} + \text{Rec}_2 \frac{1}{(1+i)^2} + \text{Rec}_3 \frac{1}{(1+i)^3} + \cdots \right) =$$

$$- i \, \text{NCA}_{\text{NPL}}^{\text{BOP}}$$

because $\text{Rec}_1 \frac{1}{(1+i)} + \text{Rec}_2 \frac{1}{(1+i)^2} + \text{Rec}_3 \frac{1}{(1+i)^3} + \cdots = \text{NCA}_{\text{NPL}}^{\text{BOP}}$ and

$\text{GCA}_{\text{oldNPL}}^{\text{EOP}} = \text{GCA}_{\text{NPL}}^{\text{BOP}} - \text{Rec}_1 - \text{wo}_{\text{old NPL}}$ (no interest accrual is applied to GCA).    q.e.d.

**Corollary B 2**: NPL Dashboard = 0 under the assumption of an unbiased cash flow estimate (ie an unbiased LGD estimate) for non-performing loans under discounting conditions of the final standard, but unwinding of interest applied to NCA and GCA simultaneously.

**Proof of Corollary B 2**:

In line with the proof in Lemma B 1, applying $\text{GCA}_{\text{oldNPL}}^{\text{EOP}} = \text{GCA}_{\text{NPL}}^{\text{BOP}} - \text{Rec}_1 + i \, \text{NCA}_{\text{NPL}}^{\text{BOP}} - \text{wo}_{\text{old NPL}}$ instead of $\text{GCA}_{\text{oldNPL}}^{\text{EOP}} = \text{GCA}_{\text{NPL}}^{\text{BOP}} - \text{Rec}_1 - \text{wo}_{\text{old NPL}}$.    q.e.d.

**Lemma B 3**: NCA = $\text{NCA}_{\text{Coll}}$ + $\text{NCA}_{\text{Unsec}}$ + $\text{NCA}_{\text{G'tee}}$ with the definitions from Section 7.

**Proof of Lemma B 3**: GCA − $\text{NCA}_{\text{Coll}}$ − $\text{NCA}_{\text{Unsec}}$ − $\text{NCA}_{\text{G'tee}}$ =

= GCA − $\text{NCA}_{\text{Coll}}$ − (1 − $\text{LGD}_u$) (GCA − $\text{NCA}_{\text{Coll}}$) − (1 − PD) $\text{LGD}_u$ (GCA − $\text{NCA}_{\text{Coll}}$) =

= $\text{LGD}_u$ (GCA − $\text{NCA}_{\text{Coll}}$) − (1 − PD) $\text{LGD}_u$ (GCA − $\text{NCA}_{\text{Coll}}$) =

= PD $\text{LGD}_u$ (GCA − $\text{NCA}_{\text{Coll}}$) = EL = GCA − NCA, or

$\text{NCA}_{\text{Coll}}$ + $\text{NCA}_{\text{Unsec}}$ + $\text{NCA}_{\text{G'tee}}$ = NCA    q.e.d.

Expected Loss Estimation for IFRS 9              V3, forthcoming in Credit Technology 92, 9/2015**References**

Benston G. J. and Wall L. D. (2005): How Should Banks Account for Loan Losses. Economic Review Fourth Quarter 2005, 19-38

Berd, Arthur M. (2013): Lessons from the Financial Crisis 2nd Impression, Insights from the defining economic event of our lifetime. Incisive Media, Risk Books

BIS (2013a): Directive 2013/36/EU of the European parliament and of the council of 26 June 2013

BIS (2013b): Regulation (EU) No 575/2013 of the European parliament and of the council of 26 June 2013

Böcker, Klaus (2010): Rethinking Risk Measurement and Reporting Vol II. Incisive Media, Risk Books

Diebold, Francis X.; Doherty, Neil A.; Herring, Richard J. (2010): The Known, The Unknown, The Unknowable in Financial Risk Management, Princeton

Domikowsky Christian, Bornemann Sven, Düllmann Klaus, Pfingsten Andreas (2014): Loan Loss Provisioning and Procyclicality: Evidence from (more than) an Expected Loss Model, Working paper Sept 9, 2014

EBA (2013): Third interim report on the consistency of risk-weighted assets: SME and residential mortgages, Dec 17, 2013

EBA (2014a): Consultation Paper, Draft Regulatory Technical Standards on materiality threshold of credit obligation past due, Oct 31, 2014

EBA (2014b): Consultation Paper, Draft Regulatory Technical Standards on the specification of the assessment methodology for competent authorities regarding compliance of an institution with the requirements to use the IRB Approach, Nov 12, 2014

Gaber, Thomas (2013): Die Qualität der Finanzberichterstattung bei Banken: Empirische Untersuchungen zu Ergebnisqualität und Wertrelevanz, Dissertation, Graz, August 2013

Grünberger, David (2012): Expected Loan Loss Provisions, Business- and Credit Cycles (December 10, 2012). Available at SSRN: http://ssrn.com/abstract=2187515.

Grünberger, David (2013): Kreditrisiko im IFRS-Abschluss: Handbuch für Bilanzersteller, Prüfer und Analysten, Schäffer/Poeschel
Wolfgang Reitgruber                                                                                                                   Page 30